\documentclass[sigconf]{acmart}

\usepackage{rotating}
\usepackage{pifont}
\usepackage{array}
\usepackage{multirow}
\usepackage{enumitem}
\usepackage{makecell}

\def\pname{{RSSD}}

\newcommand{\cmark}{\ding{52}}
\newcommand{\xmark}{\ding{55}}
\newcommand{\hrecover}{\ding{119}}
\newcommand{\frecover}{\ding{108}}
\newcommand{\nrecover}{\ding{109}}
\usepackage{def}

\usepackage{setspace}
\begin{document}

\title{\pname{}: Defend against Ransomware with Hardware-Isolated Network-Storage Codesign and Post-Attack Analysis*}

\date{}
\author{Benjamin Reidys}
\affiliation{%
    \country{UIUC}
}
\author{Peng Liu}
\affiliation{%
    \country{Pennsylvania State University}
}
\author{Jian Huang}
\affiliation{%
    \country{UIUC}
}

\authornote{This work has been published at ASPLOS'22~\cite{rssd}}
\settopmatter{printfolios=true}
\pagenumbering{arabic}

\begin{abstract}
Encryption ransomware has become a notorious malware. It encrypts user data 
on storage devices like solid-state drives (SSDs) and demands a ransom to restore data for users. 
To bypass existing defenses, ransomware would keep evolving and performing new attack models. 
For instance, we identify and validate three new attacks, including  
(1) garbage-collection (GC) attack that exploits 
storage capacity and keeps writing data to trigger GC and 
force SSDs to release the retained data; (2) timing attack that intentionally 
slows down the pace of encrypting data and hides its I/O patterns to escape 
existing defense; (3) trimming attack that utilizes the trim 
command available in SSDs to physically erase data. 

To enhance the robustness of SSDs against these attacks, 
we propose \pname{}, a ransomware-aware SSD. 
It redesigns the flash management of SSDs for enabling the hardware-assisted logging, 
which can conservatively retain older versions of user data and received storage operations 
in time order with low overhead. 
It also employs hardware-isolated NVMe over Ethernet to expand local storage 
capacity by transparently offloading the logs to remote cloud/servers in a secure manner. 
\pname{} enables post-attack analysis by building a trusted evidence chain 
of storage operations to assist the investigation of ransomware attacks. 
We develop \pname{} with a real-world SSD FPGA board. 
Our evaluation shows that    
\pname{} can defend against new and future ransomware attacks, while introducing 
negligible performance overhead.  
\end{abstract}
\maketitle

\section{Background and Motivation}
\label{sec:intro}
Although secure storage systems have been developed for decades, 
encryption ransomware imposes new challenges and has become one of the biggest cybersecurity 
threats. 
It stealthily encrypts user data and demands ransom 
from users to restore their data. 
Recent studies report that ransomware attack could happen every 11 seconds,   
the victims include 
governments, 
schools, hospitals, police departments, and personal computers.
Each attack requests an average of \$8,100 and costs nearly \$300,000 in server downtime.
These ongoing ransomware outbreaks and global damage reflect the fact that 
the current security design of storage systems falls short of defending 
against encryption ransomware.

\begin{table}[t]
\centering
        \caption{
	Comparison with state-of-the-art approaches. The 6th column represents Data Recovery (\nrecover: Unrecoverable, \hrecover: Partially Recoverable, \frecover: Recoverable). The 7th column represents Post-Attack Analysis or Storage Forensics.  
        }
\vspace{-2ex}
\label{tab:comparison}
\scriptsize
        \begin{tabular}{|p{2pt}<{\centering}|p{45pt}<{\centering}|p{10pt}<{\raggedleft}|p{20pt}<{\centering}|p{25pt}<{\centering}|p{23pt}<{\centering}|p{23pt}<{\centering}|}
\hline
               \multirow{2}{*}[0em]{}  & \multirow{2}{*}[0em]{\textbf{Related}} & \multicolumn{3}{c|}{\textbf{Defend New Attacks}} & \multirow{2}{*}[0em]{\textbf{Recovery}} & \multirow{2}{*}[0em]{\textbf{Forensics}} \\
		\cline{3-5}
		& & \textbf{GC} & \textbf{Timing} & \textbf{Trimming} & & \\
\hline
                \multirow{5}{*}[0em]{\begin{turn}{90}Software\end{turn}} & Unveil~\cite{usenix2016unveil} & \xmark & \xmark & \xmark & \nrecover & \xmark  \\
                 & CryptoDrop   & \xmark & \xmark & \xmark & \nrecover & \xmark  \\
                  & CloudBackup                        & \xmark & \cmark & \xmark & \hrecover & \xmark  \\
                  & ShieldFS                        & \xmark & \xmark & \xmark & \hrecover & \xmark  \\
                  & JFS                         & \xmark & \xmark & \xmark & \nrecover & \xmark  \\
\hline
                \multirow{4}{*}[0em]{\begin{turn}{90}Hardware\end{turn}} & FlashGuard~\cite{jian:flashguard} & \cmark & \xmark & \xmark & \hrecover & \xmark  \\
                 & TimeSSD   & \cmark & \xmark & \xmark & \hrecover & \xmark  \\
                  & SSDInsider                        & \xmark & \xmark & \xmark & \hrecover & \xmark  \\
                  & RBlocker                        & \xmark & \xmark & \xmark & \hrecover & \xmark  \\
\hline
                \multicolumn{2}{|c|}{\textbf{\pname{}}}   & \cmark & \cmark & \cmark & \frecover & \cmark \\
\hline
\end{tabular}
\vspace{-2ex}
\end{table}
To defend against ransomware attacks, software-based approaches, such as intrusion 
detection and data backup have been proposed.
Unfortunately, software-based solutions suffer from four major limitations.
First, since software-based solutions are not hardware isolated from malicious processes, they can be compromised by ransomware.
Particularly, attackers could obtain OS kernel privilege and terminate software-based backup systems.
Moreover, even though the ransomware detection succeeds, some files have been encrypted and victims 
still have to pay to get their data back. 
Second, attackers can hide in the system long before deploying their ransomware.
Third, ransomware can overwrite data backups with encrypted versions.
Finally, software-based solutions usually lack the capability of trusted post-attack analysis,which impedes the progress of recovering from an attack.

To defend against ransomware attacks, recent work exploited the intrinsic flash properties to detect ransomware attacks and restore 
    victim data.  However, they have three major limitations.
First, they were mainly developed to defend against existing
encryption ransomware which assumed the underlying storage
devices perform like conventional HDDs.
As SSDs have been widely used,
ransomware will evolve and update their attack models. 
Therefore, we must anticipate and proactively prevent new ransomware attacks.
Second, due to the limited storage capacity,
we can only retain the stale data for a limited time. 
This will significantly affect storage performance, especially for data-intensive workloads.
Even worse, ransomware could take advantage of limited storage capacity to initiate new attacks. 
Third, most defenses do not support post-attack or forensic analysis, 
which will miss the opportunity to learn new attack models. 
This slows down the post-attack investigation and limits their ability to adapt to 
evolving malware. 


\section{Threat Model}
\label{sec:threat}
As discussed in $\S$\ref{sec:intro}, malicious users could elevate their
privilege to run as administrators and disable/destroy the software-based data backup solutions.
We do not assume the OS is trusted, instead, we trust the SSD firmware. 
We believe this is a realistic threat model for two reasons.
First, the SSD firmware is located within the storage controller, underneath the generic
block interface. It is hardware-isolated from higher-level malicious
processes. Second, SSD firmware has a much smaller trusted computing base (TCB) than the OS kernel,
making it typically less vulnerable to malware attacks. Once the firmware is flushed
into the SSD controller, commodity SSDs will not allow firmware
modifications without authentication, which guarantees the implementation integrity.

\section{Design and Implementation}
\label{sec:design}
\textbf{Ransomware 2.0:}
As SSDs have become prevalent in a vast majority of computing platforms because of  
their increased performance and decreased cost, we believe 
new ransomware attack models with awareness of flash properties will happen, 
creating a new generation of security threats that we call Ransomware 2.0. These new attacks will both encrypt user data 
and defeat existing data protection schemes. 

In this paper, we present three new ransomware attacks that can circumvent existing SSD-based protections: 
(1) \textit{GC attack}, in which ransomware exploits the limited storage capacity of SSDs and 
dumps data to occupy the available storage space and 
force SSDs to release retained data. (2) \textit{Timing attack}, 
in which ransomware intentionally slows down the pace of encrypting data and 
hides its I/O patterns behind user operations. 
(3) \textit{Trimming attack}, in which ransomware 
utilizes the \emph{trim} command in commodity SSDs to erase 
flash pages and remove the original copies of encrypted data. 

It is not easy to defend against these new attacks, since  
each of them will generate new I/O patterns that can bypass existing 
detection and defense mechanisms. 
For instance, existing detection approaches worked 
by identifying the repeating I/O patterns.
The timing attack can bypass it by intentionally lowering the attack frequency and imitating regular 
storage I/O patterns. 
The GC attack can invalidate existing data recovery schemes  
by forcing the SSD to conduct GC operations and erase the retained data. And the trimming attack 
enables ransomware to speed up the removal of the original data copies that have been encrypted. 

\begin{figure}[t]
\centering
\includegraphics[width=0.53\linewidth]{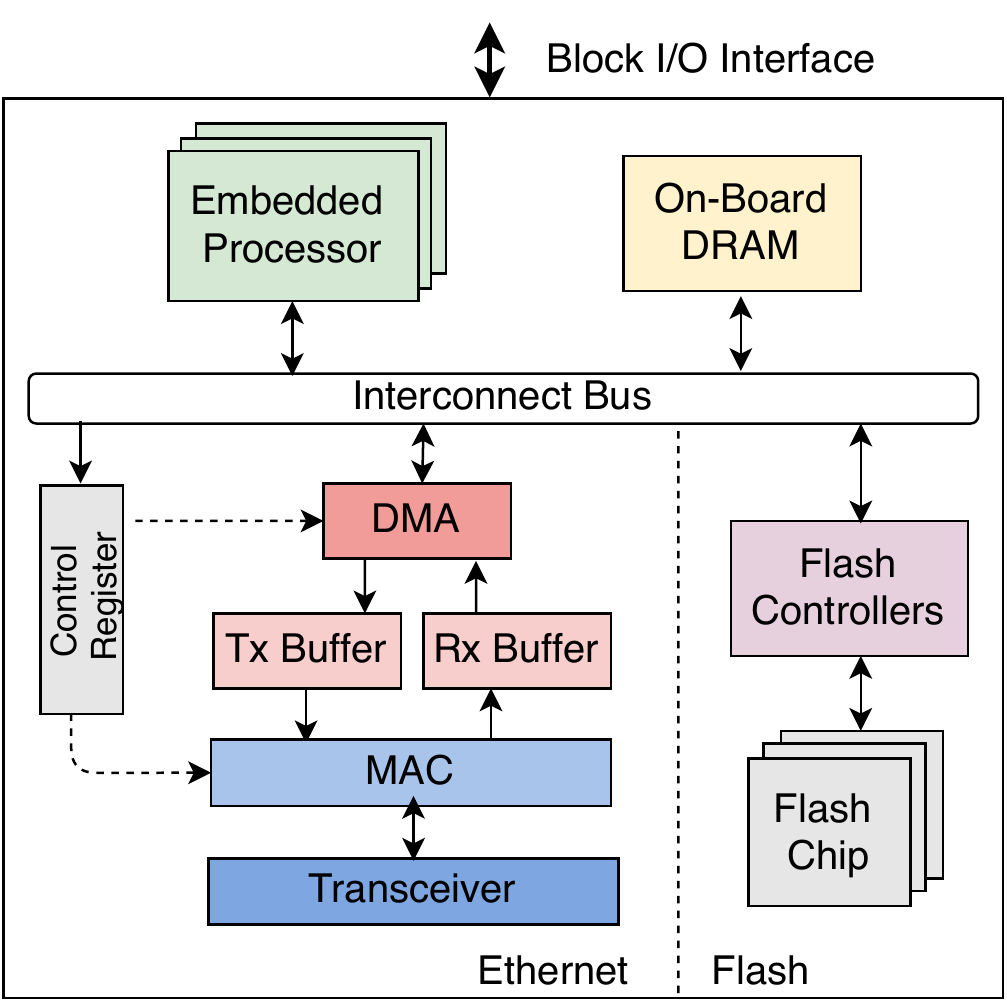}
\vspace{-3ex}
    \caption{{Architecture of \pname{}}}
\label{fig:arch}
\vspace{-3ex}
\end{figure}

From our analysis of Ransomware 2.0, we identify zero data loss recovery and trusted post-attack analysis as critical for 
solutions against evolutions of ransomware.
To implement these features, we develop a ransomware-aware SSD named \pname{}, shown in Figure~\ref{fig:arch}.  

\noindent
\textbf{Zero data loss recovery:}
\pname{} enables zero data loss recovery by conservatively retaining all stale data. 
Thus, \pname{} guarantees that all data that may be locked by ransomware is retained. 
However, this can incur significant performance overhead. Therefore, 
\pname{} proposes  a hardware-isolated NVMe over Ethernet (NVMe-oE) to transfer the retained 
pages in a compressed and encrypted format to remote cloud or storage servers in time order, 
while keeping the valid pages locally for performance. 
This enables \pname{} to expand its local storage capacity in a secure and transparent manner.

To mitigate the trimming attack, we rethink the hardware support for the \emph{trim} command in SSDs. 
The feature of using \emph{trim} commands to directly notify SSDs to garbage collect 
flash pages, would be attractive to ransomware, 
since it can bypass existing defenses. 
Instead of disabling the \emph{trim} command, \pname{} enhances it. 
Specifically, upon receiving \emph{trim} commands, \pname{} 
will allocate new flash pages and remap the addresses touched by the \emph{trim} command to 
these new pages.  
\pname{} will retain the trimmed data, 
therefore, it can still restore the victim data upon trimming attack.

\noindent
\textbf{Trusted post-attack analysis:}
To enable efficient and trusted post-attack analysis, 
we extend \pname{} and retain the log of storage operations 
in the SSD. Thus, \pname{} has the capability of reproducing 
the storage operations 
in the original order 
they were issued. As the logging approach is hardware isolated, \pname{} can build a trusted 
evidence chain for post-attack or forensic analysis. Since most of the retained logs and  
data will be transferred remotely,
\pname{} enables 
the offloading of ransomware detection and analysis to remote servers. Therefore, we can 
detect ransomware more efficiently and accurately by utilizing the powerful computing resources
and the flexibility of deploying various detection algorithms. \pname{} also allows fast 
reconstruction of evidence chains by backtracking storage operations with our 
hardware-assisted logging. 

\noindent
\textbf{Key contributions:}
We list the key contributions of this work.

\begin{itemize}[leftmargin=*]

\item We conduct an empirical study of more than a hundred ransomware cases and confirm that 
	the lack of efficient data recovery and post-attack analysis is the major weakness of 
	modern storage systems and ransomware defense solutions. 
		
\vspace{0.5ex}
\item We present a new understanding of Ransomware 2.0, discuss and validate 
	three new ransomware attacks that include GC attack, timing attack, and trimming 
	attack. 

\vspace{0.5ex}
\item We develop a new SSD that uses the hardware-isolated NVMe-oE to 
	safely extend the retention time for the hardware-assisted logging, 
	and enable the offloading of ransomware detection and analysis to the remote cloud/servers. 

\vspace{0.5ex}
\item We rethink the storage architecture support for the \emph{trim} command in SSDs 
	and enhance its security by enabling the retention of trimmed data with hardware-assisted logging.  

\vspace{0.5ex}
\item We present a hardware-isolated post-attack analysis approach that can efficiently build a trusted 
	evidence chain of storage operations that lead up to an attack.   
\end{itemize}

\noindent
\textbf{Implementation of \pname{}:}
We implement \pname{} with a Cosmos+ OpenSSD FPGA board, 
a cloud storage service Amazon S3, and local storage servers. 
We use various storage benchmarks and I/O traces, 
and ransomware samples collected from VirusTotal~\cite{VirusTot50:online}. Our evaluation shows that 
\pname{} can retain all obsolete data across the SSD and remote cloud/servers, with minimal storage cost, and 
negligible impact on the local storage performance and device lifetime. 
We replay multiple ransomware attacks on \pname{}, and show that \pname{} 
can restore the data encrypted by ransomware, and correctly reconstruct the original sequence of 
I/O events that lead to the attacks in a short time.

\begin{figure}[t]
\centering
\includegraphics[width=0.98\linewidth]{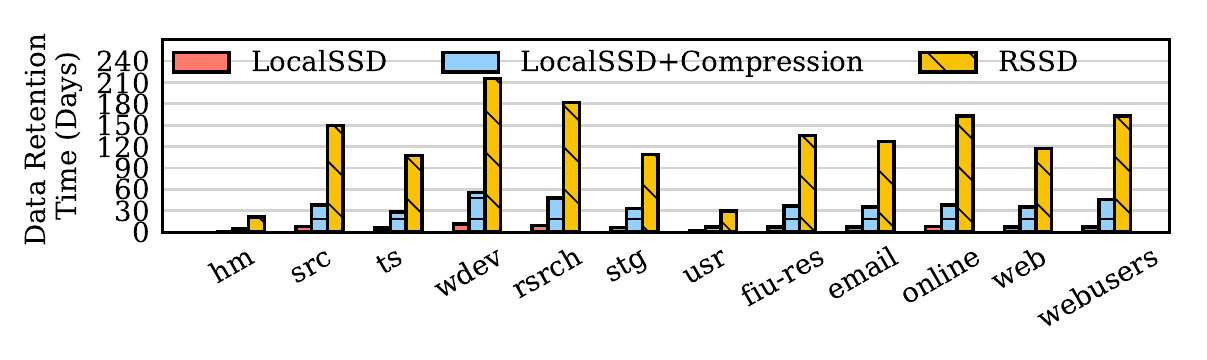}
\vspace{-4ex}
\caption{{Data retention time in \pname{}.}}
\label{fig:ret}
\vspace{-3ex}
\end{figure}

\noindent
\textbf{Performance of \pname{}:}
Our evaluation shows that (1) \pname{} can retain the stale data for a much longer
time than state-of-the-art approaches, over 200 days in our evaluation (see Figure~\ref{fig:ret});
(2) It has less than 1\% negative impact on storage
performance 
and minimal impact on device lifetime;
(3) It performs fast data recovery after attacks;
(4) It enables efficient post-attack analysis by building a trusted chain of I/O operations (see~\cite{rssd} for the full evaluation).

\bibliographystyle{ACM-Reference-Format}
\bibliography{secref,jianref,ref,flatflash,deepstore}


\begin{thebibliography}{4}


\ifx \showCODEN    \undefined \def \showCODEN     #1{\unskip}     \fi
\ifx \showDOI      \undefined \def \showDOI       #1{#1}\fi
\ifx \showISBNx    \undefined \def \showISBNx     #1{\unskip}     \fi
\ifx \showISBNxiii \undefined \def \showISBNxiii  #1{\unskip}     \fi
\ifx \showISSN     \undefined \def \showISSN      #1{\unskip}     \fi
\ifx \showLCCN     \undefined \def \showLCCN      #1{\unskip}     \fi
\ifx \shownote     \undefined \def \shownote      #1{#1}          \fi
\ifx \showarticletitle \undefined \def \showarticletitle #1{#1}   \fi
\ifx \showURL      \undefined \def \showURL       {\relax}        \fi
\providecommand\bibfield[2]{#2}
\providecommand\bibinfo[2]{#2}
\providecommand\natexlab[1]{#1}
\providecommand\showeprint[2][]{arXiv:#2}

\bibitem[\protect\citeauthoryear{Huang, Xu, Xing, Liu, and Qureshi}{Huang
  et~al\mbox{.}}{2017}]%
        {jian:flashguard}
\bibfield{author}{\bibinfo{person}{Jian Huang}, \bibinfo{person}{Jun Xu},
  \bibinfo{person}{Xinyu Xing}, \bibinfo{person}{Peng Liu}, {and}
  \bibinfo{person}{Moinuddin~K. Qureshi}.} \bibinfo{year}{2017}\natexlab{}.
\newblock \showarticletitle{{FlashGuard: Leveraging Intrinsic Flash Properties
  to Defend Against Encryption Ransomware}}. In
  \bibinfo{booktitle}{\emph{Proceedings of the 24th ACM Conference on Computer
  and Communications Security (CCS'17)}}. \bibinfo{address}{Dallas, TX}.
\newblock


\bibitem[\protect\citeauthoryear{Kharaz, Arshad, Mulliner, Robertson, and
  Kirda}{Kharaz et~al\mbox{.}}{2016}]%
        {usenix2016unveil}
\bibfield{author}{\bibinfo{person}{Amin Kharaz}, \bibinfo{person}{Sajjad
  Arshad}, \bibinfo{person}{Collin Mulliner}, \bibinfo{person}{William
  Robertson}, {and} \bibinfo{person}{Engin Kirda}.}
  \bibinfo{year}{2016}\natexlab{}.
\newblock \showarticletitle{UNVEIL: A Large-Scale, Automated Approach to
  Detecting Ransomware}. In \bibinfo{booktitle}{\emph{25th USENIX Security
  Symposium (USENIX Security 16)}}. \bibinfo{publisher}{USENIX Association},
  \bibinfo{address}{Austin, TX}, \bibinfo{pages}{757--772}.
\newblock
\showISBNx{978-1-931971-32-4}


\bibitem[\protect\citeauthoryear{Reidys, Liu, and Huang}{Reidys
  et~al\mbox{.}}{2022}]%
        {rssd}
\bibfield{author}{\bibinfo{person}{Benjamin Reidys}, \bibinfo{person}{Peng
  Liu}, {and} \bibinfo{person}{Jian Huang}.} \bibinfo{year}{2022}\natexlab{}.
\newblock \showarticletitle{RSSD: Defend against Ransomware with
  Hardware-Isolated Network-Storage Codesign and Post-Attack Analysis}. In
  \bibinfo{booktitle}{\emph{Proceedings of the 27th ACM International
  Conference on Architectural Support for Programming Languages and Operating
  Systems(ASPLOS'22)}} (Lausanne, Switzerland).
\newblock


\bibitem[\protect\citeauthoryear{{VirusTotal - Free Online Virus, Malware and
  URL Scanner}}{{VirusTotal - Free Online Virus, Malware and URL
  Scanner}}{2016}]%
        {VirusTot50:online}
\bibfield{author}{\bibinfo{person}{{VirusTotal - Free Online Virus, Malware and
  URL Scanner}}.} \bibinfo{year}{2016}\natexlab{}.
\newblock \bibinfo{howpublished}{\\\url{https://www.virustotal.com/}}.
\newblock


\end{thebibliography}

\end{document}